\documentclass[%
reprint,
superscriptaddress,
 amsmath,amssymb,
 aps,
]{revtex4-1}

\usepackage{graphicx}% Include figure files
\usepackage{dcolumn}% Align table columns on decimal point
\usepackage{bm}% bold math
\usepackage{diagbox}% bold math
\usepackage{epstopdf}

\begin{document}

\preprint{APS/123-QED}

\title{Permutation time irreversibility in sleep electroencephalograms: Dependence on sleep stage and the effect of equal values}% Force line breaks with \\

\author{Wenpo Yao}
\affiliation{State Key Laboratory of Organic Electronics and Information Displays, Institute of Advanced Materials, School of Chemistry and Life Sciences, School of Geographic and Biologic Information, Nanjing University of Posts and Telecommunications, Nanjing 210023, China}
\affiliation{Key Laboratory of Computational Neuroscience and Brain-Inspired Intelligence(Fudan University), Ministry of Education}
%


\begin{abstract}
Time irreversibility (TIR) refers to the manifestation of nonequilibrium brain activity influenced by various physiological conditions; however, the influence of sleep on electroencephalogram (EEG) TIR has not been sufficiently investigated. In this paper, a comprehensive study on permutation TIR (pTIR) of EEG data under different sleep stages is conducted. Two basic ordinal patterns (i.e., the original and amplitude permutations) are distinguished to simplify sleep EEGs, and then the influences of equal values and forbidden permutation on pTIR are elucidated. To detect pTIR of brain electric signals, 5 groups of EEGs in the awake, stages I, II, III, and rapid eye movement (REM) stages are collected from the public Polysomnographic Database in PhysioNet. Test results suggested that the pTIR of sleep EEGs significantly decreases as the sleep stage increases (p$<$0.001), with the awake and REM EEGs, demonstrating greater differences than others. Comparative analysis and numerical simulations support the importance of equal values. Distribution of equal states, a simple quantification of amplitude fluctuations, significantly increases with the sleep stage (p$<$0.001). If these equalities are ignored, incorrect probabilistic differences may arise in the forward-backward and symmetric permutations of TIR, leading to contradictory results; moreover, the ascending and descending orders for symmetric permutations also lead different outcomes in sleep EEGs. Overall, pTIR in sleep EEGs contributes to our understanding of quantitative TIR and classification of sleep EEGs.
\end{abstract}

%\pacs{05.45.Tp, 87.85.-d}


\maketitle

%\tableofcontents

\section{Introduction}
The human brain, which contains large numbers of neurons and interacts with other physiological organs \cite{Andrea2018,Bashan2012}, is a highly complex system. Brain activity exhibits evidently nonlinear, nonequilibrium properties and is influenced by various internal and external factors. Sleep is a vital physiological activity for living beings; various changes occur in the physiological systems during sleep, with brain electric activities demonstrating pronounced changes. Therefore, the complex characteristics of sleep electroencephalograms (EEGs) can help in analyzing the sleep mechanisms \cite{Zhao2019,Phan2022}. Various nonlinear methods, such as fractal approaches and detrended fluctuation analysis, have been used to explore the dynamics of EEGs during different sleep stages. Entropy measures \cite{Ma2018,Xiong2017} are widely employed to detect the complexity of sleep brain signals, such as the Shannon entropy \cite{Hou2021,Bandt2017}, sample entropy \cite{Liang2021,Wang2021}, and multiscale entropy \cite{Miskovic2019}. Among these complex characteristics, the loss of time reversibility is a manifestation of nonequilibrium brain electric activity \cite{Fang2019,Lynn2021} and has been applied to several neural pathological conditions, including epilepsy \cite{Yao2020ND,Yao2020CNS,Marti2018}, Alzheimer’s disease \cite{Martin2019}, and alcoholism \cite{Zanin2021}.

Time irreversibility (TIR) \cite{Weiss1975}, also defined as temporal asymmetry (TAS) \cite{Kelly1979}, is an important characteristic of nonequilibrium EEGs. To quantity TIR, the probabilistic differences between the forward-backward processes or symmetry vectors must be measured \cite{Weiss1975,Kelly1979,Ramsey1995}; both are nontrivial. In real-world signal processing, TIR is generally quantified by coarse-graining the time series. Probabilistic differences between up and down were introduced by Costa \cite{Costa2005,Costa2008}, Porta \cite{Porta2008}, and Ehlers et al. \cite{Guzik2006,Ehlers1998} as simplified temporal asymmetries. Lacasa et al. \cite{Lacasa2012,Lacasa2015,Flanagan2016} estimated the irreversibility considering the distinguishability between the in-out distributions of the visibility graph \cite{Lacasa2008}. The in-out difference was then employed to detect irreversible characteristics in effective interactions and particular networked connectivity \cite{Donges2013,Zou2019}. Given nonequilibrium statistical mechanics in neuronal spike trains, the network inference and couplings of asymmetric models were investigated via asynchronous update \cite{Zeng2013,Zeng2011}. Symbolic TIR based on a way of coarse-graining or reduction of description is widely adopted in quantitative TIR owing to its computational effectiveness and simplified statistical analysis \cite{Daw2003,Cammarota2007}. Among these coarse-graining methods, ordinal patterns convey structural dynamics and do not impose further model assumptions \cite{Bandt2002,Bandt2016,Bandt2020,Yao2022Perm}; thus, they are popular in the quantification of TIR \cite{Yao2020ND,Yao2020CNS,Marti2018,Martin2019,Zanin2021}. However, several challenges are observed in permutation TIR (pTIR). It should be noted that there are two basic ordinal patterns, i.e., original permutation and amplitude permutation; they differently reflect the structural dynamics of any series \cite{Yao2022Perm}. Amplitude permutation directly reflects the vector temporal structure, while the application of original permutation might result in conceptual errors in pTIR \cite{Yao2020CNS,Yao2019Sym,Yao2023CNS}. Forbidden permutation refers to the missing ordinal pattern in the simplified transformation of a temporal structure. Amigo et al. \cite{Amigo2006,Amigo2007,Amigo2008} extensively analyzed the features of these forbidden ordinal patterns and proposed several methods for chaotic determination in real-world time series analysis. The distributions of forbidden permutation convey important system information, such as chaotic dynamics, correlation, and nonlinearity \cite{Zanin2008,Carpi2010,Kulp2017,David2020}. Forbidden permutation is an adverse factor in pTIR because it generates individual permutations and make division-based parameters (e.g., the Kullback-Leibler distance) unsuitable to calculate the probabilistic difference between forbidden and individual permutations \cite{Yao2020ND,Yao2020CNS,Yao2019Sym,Yao2023CNS}. Equal value is another implicated factor because it significantly affects the construction of ordinal patterns and permutation analysis. Ma et al. \cite{Bian2012} focused on the indexes of equal values in permutation and modified them into the same symbol (rank) for a more accurate characterization of the system structure. Zunino et al. \cite{Zunino2017} reported erroneous conclusions on permutation entropy (PEn) in the event of equalities in time series; David et al. \cite{Cuesta2018} further identified the weakness of PEn, i.e., the possible ambiguities introduced by equal values in the subsequences. However, the influence of equal values on pTIR has not been sufficiently analyzed. Equal values are generally assumed to be rare if a process has a continuous distribution \cite{Bandt2002}. This is partly true; however, equal values are observed in physiological signals such as heartbeats \cite{Yao2019E,Yao2020APL} and raw EEGs \cite{Yao2021DES,Yao2023tDES}, and they significantly impact permutation analysis. Moreover, equal values might generate self-symmetry vectors containing important physical implication (i.e., time reversibility or temporal symmetry) in TIR and produce contradictory findings in real-world series analysis \cite{Yao2019E,Yao2020APL}. Furthermore, the different treatment of equal values might lead to inconsistencies between forward-backward and symmetric permutations, thus yielding differences in the quantification of TIR and TAS of time series. Therefore, the effects on the pTIR during signal processing, particular the TIR and TAS in real-world series analysis, should be comprehensively studied. Owing to these unfavorable factors in the quantification of TIR, the time irreversible characteristics of sleep EEGs have not been given the deserved attention.

To address this problem, a comprehensive analysis of pTIR and its application in sleep EEG classification is performed. Accordingly, two basic ordinal patterns are compared to simplify the time series, and several crucial factors in pTIR are clarified. Subsequently, the distributions of equal values and individual permutations that affected the pTIR of sleep EEGs are detected. For comparison, nonequal permutations of TIR are employed, and PEn is applied to evaluate the complexity of sleep EEGs. Furthermore, several issues are discussed such as equal values in sleep EEGs and the relationship between entropy and TIR. The contributions of this research are as follows: (1) it elucidates the key factors of pTIR in physiological signal processing, especially the influence of equal values, (2) it explores the time irreversible characteristics of EEGs under different sleep conditions, which is helpful for sleep classification.

\section{Methods}

\subsection{Time irreversibility}
A process is defined as time reversible if it is invariant under the reversal of the timescale; otherwise, it is time irreversible. Given below are two statistical definitions of time reversibility:

Definition 1. According to Weiss \cite{Weiss1975}, a stationary process, $X(t)$, is time reversible if $\{X(t_{1}),X(t_{2}), \ldots,X(t_{m})\}$ and $\{X(-t_{1}),X(-t_{2}), \ldots,X(-t_{m})\}$ have the same joint probability distributions for every $t_{1},t_{2}, \ldots,t_{m}$ and $m$; otherwise, $X(t)$ is time irreversible.

Definition 2. Based on the works of Kell \cite{Kelly1979}, if $X(t)$ is time reversible, $\{X(t_{1}),X(t_{2}), \ldots,X(t_{m})\}$ and $\{X(-t_{1}+n),X(-t_{2}+n), \ldots,X(-t_{m}+n)\}$ have the same probability distributions for every $t_{1},t_{2}, \ldots,t_{m}$ and $n$. Particularly, under $n=t_{1}+t_{m}$, symmetric $\{X(t_{1}),X(t_{2}), \ldots,X(t_{m})\}$ and $\{X(t_{m}), \ldots,X(t_{2}),X(t_{1})\}$ have the same joint probabilities. Moreover, the symmetric form of a vector is the same as its counterpart in the time-reversal series. Therefore, TIR is also defined as TAS.

To quantify the TIR or TAS of a process, the forward-backward probabilistic difference and symmetric vectors' probabilistic divergence should be equivalent \cite{Ramsey1995}; however, their applications in real-world processes are different \cite{Yao2020ND,Yao2020CNS}. The forward-backward approach for TIR is operationally convenient and more reliable in application. Meanwhile, it should have the entire process to obtain the reversed one, which is not feasible if the process is large or uninterrupted; therefore, TIR based on forward-backward differences does not satisfy real-time requirement. By contrast, TAS based on the probabilistic difference of symmetric vectors demonstrates real-time performance; therefore, it has higher applicability in physiological and environmental condition monitoring. Note that TAS based on symmetric vectors has a limitation, i.e., the vector should be faithfully associated with its alternative; otherwise, conceptual misleading may occur \cite{Yao2020CNS,Yao2019Sym,Yao2023CNS}.

\subsection{Original and amplitude permutations}
TIR can be measured by alternatively calculating the joint probabilistic differences of simplified processes instead of the raw process. Among these simplified measures, pTIR is particularly popular. The ordinal pattern comes naturally from the series and does not pose further model assumptions \cite{Bandt2002,Yao2022Perm}; hence, it plays an important role in quantitative TIR.

According to the representation of a vector structure, there are two basic ordinal patterns, i.e., the original permutation (OrP) and amplitude permutation (AmP) \cite{Yao2022Perm,Yao2023CNS}. The OrP consists of the indexes of reorganized values in the original series, while the AmP comprises the positions of the original values in the reordered series. Given series $X(i)=\{x(i_{1}),\ldots,x(i_{i}),\ldots,x(i_{m})\}$, it is reordered in the ascending or descending order to $X(j)=\{x(j_{1}),\ldots,x(j_{j}),\ldots,x(j_{m})\}$ e.g., increased as $x(j_{1})<\cdots < x(j_{j})<\cdots <x(j_{m})$. Then, the OrP and AmP are generated according to the organized indexes of the original and reordered series as Eq.~(\ref{eq1}).

\begin{eqnarray}
	\label{eq1}
	\left\{
	\begin{array}{lr}
		\mbox{OrP}:(x_{1,j},\ldots,x_{i-1,j},x_{i,j},x_{i+1,j},\ldots,x_{m,j})\\
		\mbox{AmP}: (x_{i,1},\ldots,x_{i,j-1},x_{i,j},x_{i,j+1},\ldots,x_{i,m})
	\end{array}
	\right.
\end{eqnarray}

In the construction of OrP, $i$ increases from 1 to $m$ and $j$ is the location of reorganized values in the original series $X(i)$, i.e., $OrP_{j}=(j_{1},j_{2},\ldots,j_{i-1},j_{i},j_{i+1},\cdots, j_{m-1},j_{m})$. In the generation of AmP, $j$ increases from 1 to $m$ and $i$ is the position of original values in the reordered series $X(j)$, i.e., $AmP_{i}=(i_{1},i_{2},\ldots,i_{j-1},i_{j},i_{j+1},\cdots, i_{m-1},i_{m})$. Taking a series with five values $X(i)=\{5,1,7,3,9\}$ as an example, the reordered series can be represented as $X(j)=\{1,3,5,7,9\}$ in the ascending order. The indexes of reorganized $X(j)$ values in the original $X(i)$ is OrP=(2,4,1,3,5); the positions of the original $X(i)$ values in the reorganized $X(j)$ is AmP=(3,1,4,2,5).

\begin{figure*}[htb]
	\centering
	\includegraphics[width=16cm,height=5cm]{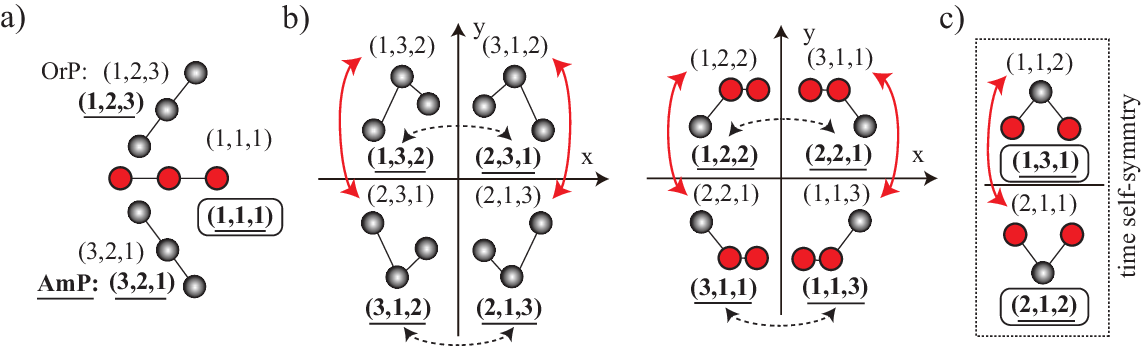}
	\caption{OrP and AmP of triple-value series. AmPs are bold and underlined. Indexes of red equal values in the ascending order are modified to be the smallest ones in their corresponding groups. a) OrPs and AmPs of all-up, all-down, and all-equal vectors are always the same. b) Symmetric AmPs of time-symmetric ($y$-axis) series are connected by dashed black arrows; symmetric OrPs of $x$-axis symmetric series are connected by solid red arrows. c) AmPs of time self-symmetry vectors are symmetric.}
	\label{fig1}
\end{figure*}

Equal values are not rare in real-world signals and are usually generated owing to limitations in signal collection \cite{Yao2021DES,Yao2023tDES}, especially the quantization error in analog-to-digital conversion (ADC). Equal values have an important role in the construction of ordinal patterns and permutation analysis \cite{Bian2012,Zunino2017,Cuesta2018,Yao2019E,Yao2020APL}; therefore, their indexes in ordinal patterns should be improved accordingly. If there are equal values in series $X(i)$, they can be organized in neighboring orders according to their order of occurrence; for example, $\cdots<x(i_{1},j_{1})=x(i_{2},j_{2})< \cdots <x(i_{3},j_{3})=x(i_{4},j_{4})=x(i_{5},j_{5})<\cdots$. Then, the indexes of equal values can be rewritten to be the same in each group, such as to the smallest indexes as $\cdots<x(i_{1},j_{1})=x(i_{1},j_{1})< \cdots <x(i_{3},j_{3})=x(i_{3},j_{3})=x(i_{3},j_{3})<\cdots$ \cite{Bian2012} or the largest ones as $\cdots<x(i_{2},j_{2})=x(i_{2},j_{2})< \cdots <x(i_{5},j_{5})=x(i_{5},j_{5})=x(i_{5},j_{5})<\cdots$, and modify the OrP and AmP accordingly. Further taking a series with five values $X(i)=\{5,1,9,1,7\}$ as an example, the second `1' could be treated as `2' according to its order of occurrence; then, the OrP and AmP become (2,4,1,5,3) and (3,1,5,2,4) in the ascending order, respectively. In equal-value ordinal patterns, the indexes of equal values are modified to be the smallest; consequently, the OrP and AmP are improved to be (2,2,1,5,3) and (3,1,5,1,4), respectively. Without equal values, if the length of the series is $m$, there exist $m$! ordinal patterns, and there are more motifs if the equality is considered. Equal-value permutation is necessary for the comprehensive reflection of the series’ temporal structure \cite{Yao2022Perm,Yao2023CNS}. Figure~\ref{fig1} illustrates the structures of triple-value series and their OrPs and AmPs.

Figure~\ref{fig1} displays the comprehensive temporal structures of triple-value series where equal values have an important role. Furthermore, OrP and AmP are two basic ordinal patterns that differently convey the temporal structure of the series \cite{Yao2022Perm}. AmP directly reflects the temporal structure of the series because its elements directly correspond to the amplitude of the original sequence element. Therefore, AmPs of symmetric vectors (e.g., $y$-axis or time symmetric) are all symmetric in Fig.~\ref{fig1}, and AmPs of the three temporally self-symmetric vectors are also self-symmetric. By contrast, OrP indirectly reflects the original series structure owing to targets on the reordered series, and OrPs of amplitude-symmetric (i.e., $x$-axis symmetry) series are symmetric. The differences between OrP and AmP differently affect permutation analysis. When ordinal patterns are used as an alternative label or symbol of the series, there are no differences between the two basic permutations, such as in PEn \cite{Bandt2017,Bandt2002}. Otherwise, if ordinal patterns are constructed as a direct replacement of the series structure, OrP and AmP should be dealt with carefully to avoid possible errors, such as in pTIR \cite{Yao2020CNS,Yao2019Sym,Yao2023CNS}. In this study, AmP is used as a direct alternative for the temporal structure of the series to avoid possible errors in pTIR. 

\subsection{Permutation time irreversibility}
Permutation TIR is the probabilistic difference between forward-backward or symmetry permutations as an alternative to the original series. Owing to the advantages of ordinal patterns and their application in quantitative TIR, pTIR is widely used in time series analysis. In real world signal analysis, the application of pTIR encounters some challenges, as listed below:

(1) OrP is not a direct reflection of the temporal structure of vectors, as shown in Fig.~\ref{fig1}, and its effect on TAS is such that if the probabilistic difference between symmetric vectors is calculated, symmetric OrPs should not be used. The AmPs of time-symmetric vectors are symmetric, and symmetric AmPs could be employed as alternatives for evaluating TIR \cite{Yao2022Perm}. Meanwhile, the OrPs of amplitude-symmetric vectors are symmetric, and symmetric OrPs should be employed as alternatives for measuring amplitude irreversibility \cite{Yao2020CNS,Yao2023CNS}. It should be noted that when measuring the joint probabilistic difference of the forward-backward process for TIR, there is no difference between the OrP and AmP. Overall, AmP directly reflects the temporal structure of vectors and is recommended for pTIR.

(2) The consideration of equal values is necessary to construct comprehensive and reliable vector structures; more importantly, equal values can generate self-symmetric vectors. Self-symmetric vectors, e.g., the vectors and their AmPs in boxes in Fig.~\ref{fig1}a) and c), have a special physical implication, i.e. time reversibility or temporal symmetry \cite{Yao2022Perm,Yao2023CNS}. Furthermore, equal values widely exist in physiological data (either directly collected EEGs or indirect heartbeats derived from electrocardiography) owing to the nonlinear and irreversible quantization process of signal collection \cite{Yao2019E,Yao2021DES,Yao2023tDES}. Therefore, equal values should not be broken by adding small random perturbations or ranked according to their order of appearance, and their indexes in ordinal patterns should be modified accordingly. 

(3) Forbidden permutations are a special type of forbidden symbol, i.e., symbols that do not exist for a process. Forbidden permutations are closely related to system characteristics \cite{Zanin2008,Carpi2010,Kulp2017,David2020,Amigo2006,Amigo2007,Amigo2008}, but they negatively impact quantitative TIR. Among the pairs of permutations for quantitative TIR, if there exists a forbidden permutation, its corresponding permutation is an individual permutation. The mathematical difference between the probabilities of forbidden and individual permutations is zero or infinite if using division-based parameters, such as the Kullback-Leibler distance. Therefore, such division-based parameters are not suitable for quantitative TIR \cite{Yao2020ND,Yao2020CNS,Yao2019Sym,Yao2023CNS}. It is recommended to use subtraction-based parameters, such as $Y_{s}$ in Eq.~(\ref{eq2}), in quantitative TIR to calculate the probabilistic differences, where $p_{i}$ and $p_{j}$ probability of corresponding permutations and $p_{i}$ should not be less than $p_{j}$.

\begin{eqnarray}
	\label{eq2}
	Y_{s} \langle p_{i},p_{j} \rangle = p_{i} \frac{p_{i}-p_{j}}{p_{i}+p_{j}},
\end{eqnarray}

\begin{figure}[htb]
	\centering
	\includegraphics[width=6.9cm,height=5.5cm]{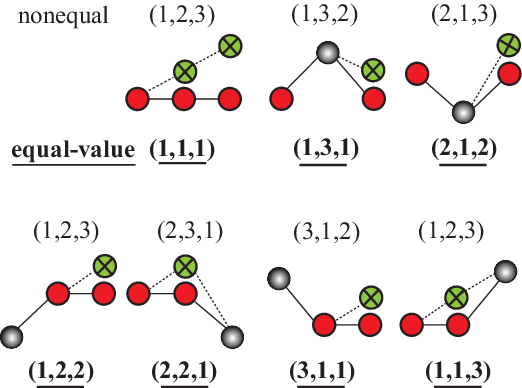}
	\caption{AmPs of triple-value series considering equalities. Equal values are represented in red and the crossed greens are their alternatives according to their order of occurrence in the ascending order. In equal-value AmPs (bold and underlined), indexes of red equal values are modified to be the smallest in their corresponding groups.}
	\label{fig2}
\end{figure}

(4) TIR and TAS are statistically consistent; theoretically, the probabilistic differences between symmetric permutations and those between forward-backward permutations should be the same. However, the existence of equal values and the traditional treatment that ranks them according to their order of emergence make TIR and TAS different. Figure~\ref{fig2} illustrates a comparative construction of nonequal and equal-value AmPs considering equal values.

As evident in Fig.~\ref{fig2}, nonequal AmP represents a temporal structure of vector without equal values. Equal values are transformed into `false up' in nonequal ordinal patterns and their symmetric permutation is `real down’, but their corresponding permutations in the backward series are `false up' and `real up'. Note that `real down' in the forward series is `real up' in the backward series; moreover, the probability distributions of `up' in the forward and backward series are both incorrect. Furthermore, if there are equal values in the process and nonequal permutation is applied, neither TIR nor TAS can be correctly quantified; in addition, the AmPs of symmetric and forward-backward vectors are not the same, leading to inconsistent and even contradictory TIR and TAS in the quantitative nonequilibrium of signal processing, which will be confirmed in this study.

\section{Results}
EEG is a typically complex signal and subject to different physiological activities. In this section, EEG data under different sleep conditions are collected from the publicly available PhysioNet \cite{Goldber2000} to test the sleep stages on pTIR. The probabilistic differences of AmPs are calculated using the subtraction-based $Y_{s}$ to quantify TIR; TIR and TAS based on equal-value AmP are denoted as pTIR and pTAS, while those based on nonequal AmP are represented as noeTIR and noeTAS, respectively.

\subsection{Sleep EEGs}
The MIT-BIH Polysomnographic Database is a collection of sleep physiologic data. This database contains information related to the sleep physiological signals of 16 male subjects (age ranging from 32 to 56, mean age 43; weight ranging from 89 to 152 kg, mean weight 119 kg). The database contains over 80 h of four-, six-, and seven-channel polysomnographic recordings, with a standard expert annotation for sleep stages after every 30 s according to the criteria of Rechtschaffen and Kales \cite{Rechtsch1968}. EEG signals are recorded from the C4-A1, O2-A1, or C3-O1 channel at a sampling rate of 250 Hz and 12-bit quantization. Referring to the annotation files, sleep EEGs in five stages, i.e., awake and sleep stages 1 (SI), 2 (SII), 3 (SIII), and rapid eye movement (REM) are extracted; each stage contained 45 sets of EEG signals with a duration of 60 s (15000 points) after visual inspection for artifacts. More detailed information can be found in Ref \cite{Ichima1999,Goldber2000}. The nonparametric Mann--Whitney U test is performed to test the statistical differences in pTIR between each of the two stages of sleep EEGs, and Kruskal--Wallis analysis of variance is applied to measure those in pTIR of sleep EEGs in five stages.

\begin{figure*}[htb]
	\centering
	\includegraphics[width=17cm,height=5.9cm]{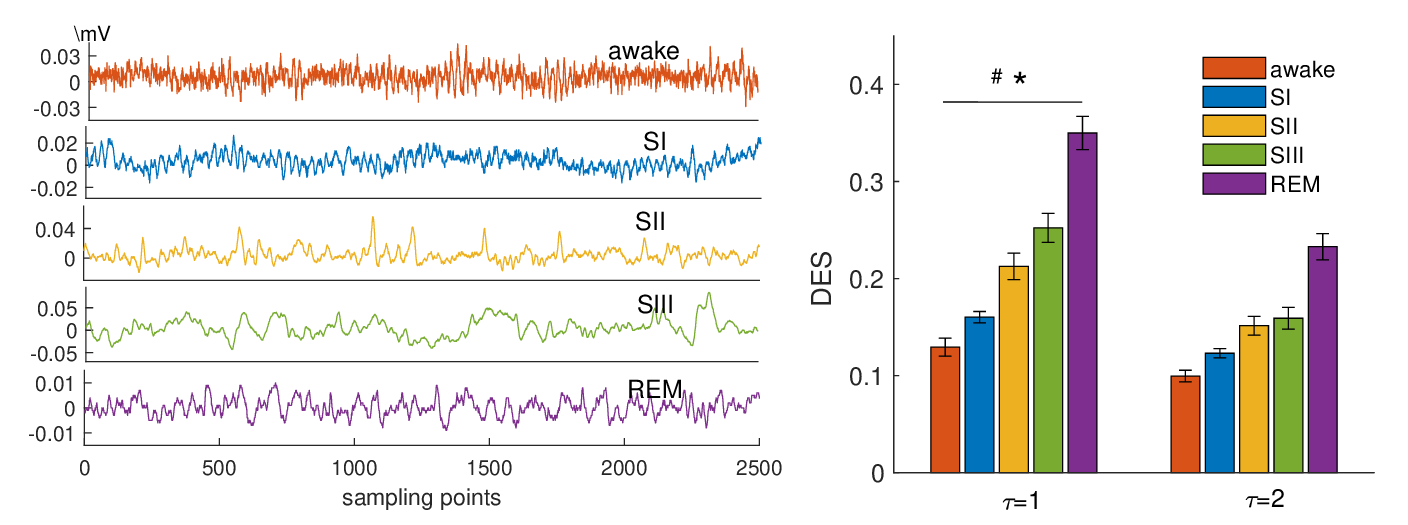}
	\caption{Exemplary EEGs in awake, SI–SIII, and REM states and the distribution of equal states (mean$\pm$standard error) of sleep EEGs. The state in DES is the direct amplitude value during sleep EEGs. \# indicates p$<$0.0001 across all stages using the Kruskal--Wallis test and * suggests p$<$0.01 between each two stages of sleep EEGs using Mann--Whitney U test.}
	\label{fig3}
\end{figure*}

Equal values play an important role in the construction of ordinal patterns and might significantly change the probability distribution of permutations; moreover, they convey important physical implication, i.e., time reversibility \cite{Yao2020ND,Yao2020CNS,Yao2022Perm,Yao2023CNS}. The distribution of equal states (DES) \cite{Yao2021DES} is measured by Eq.~(\ref{eq3}), where $L$ denotes the length of series, $N(s(t)=s(t+\tau))$ represents the number of neighboring equal states with delay $\tau$. The states of DES indicate EEG values in this report. DES of five groups of sleep EEGs are shown in Fig.~\ref{fig3}.

\begin{eqnarray}
	\label{eq3}
	\mbox{DES} = \frac{N(s(t)=s(t+\tau))}{L-\tau}
\end{eqnarray}

The results in Fig.~\ref{fig3} are consistent with those obtained in our previous report on the distribution of equal values in sleep EEGs \cite{Yao2023tDES}. Raw sleep EEGs have numerous neighboring equal values. Although awake EEGs have minimum equal values, their DES are 12\%, and as the sleep stages increase, the DES increase significantly; hence, REM EEGs have almost 34\% equal values. Under acceptable data recording resolution, awake EEGs have larger amplitude fluctuations that produce fewer neighboring equal values, and as the depth of sleep increases, amplitude fluctuations of the brain's electrical activity decrease, thus significantly decreasing the DES. Equal values are generated owing to the limitation of the ADC, i.e., the zero-amplitude fluctuation \cite{Yao2021DES,Yao2023tDES}. Test results suggest that the DES of EEGs under the five sleep conditions are statistically significantly different (p$<$0.01 for Mann--Whitney U test; p$<$0.0001 for Kruskal--Wallis test), and the DES of REM EEGs are particularly different between others (p$<$0.00001). Therefore, equal values not only have important effects on pTIR, but their distribution also serves as a simple parameter for time-domain feature extraction and should not be ignored.

\begin{figure*}[htb]
	\centering
	\includegraphics[width=17cm,height=5.7cm]{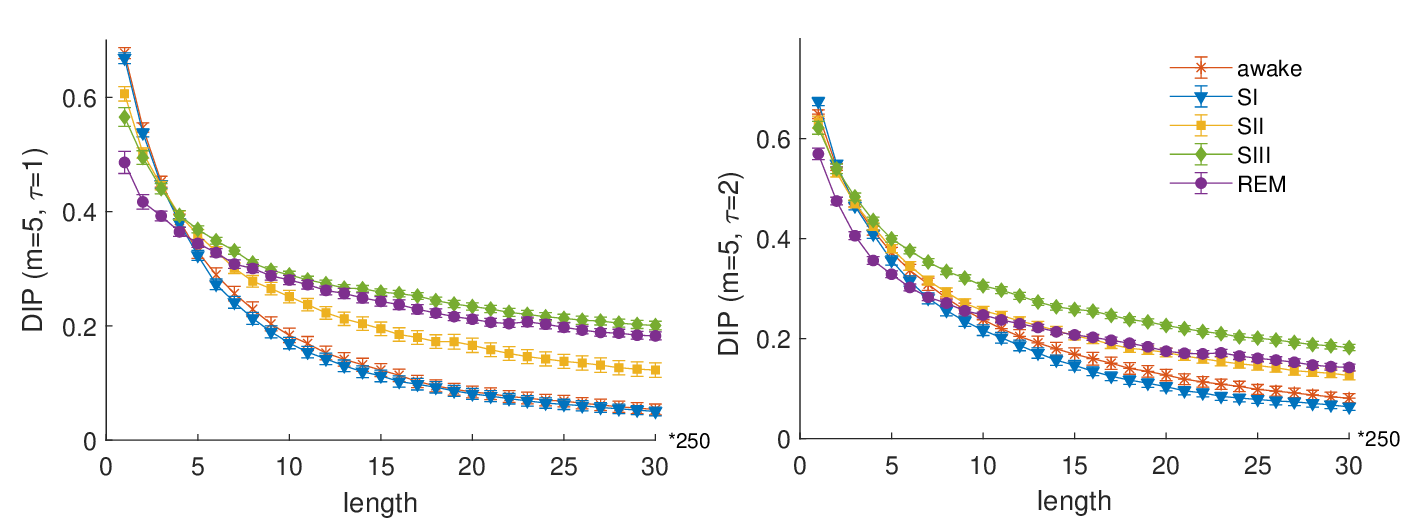}
	\caption{Distribution of individual permutations (mean$\pm$standard error) of wake, SI–SIII, and REM EEGs.}
	\label{fig4}
\end{figure*}

The existence of forbidden permutations makes division-based parameters unsuitable for pTIR, and their distribution has been widely proved to be closely associated with systematic information \cite{Zanin2008,Carpi2010,Kulp2017,David2020,Amigo2006,Amigo2007,Amigo2008}. In sleep EEGs, all permutations have their corresponding forms when $m$=2 and 3, while the distribution of individual permutations is rare when $m$=4. When $m$ is 5 and larger, there exist forbidden as well as individual permutations. Further, forbidden permutations contain false forbidden permutations that do not exist because the data length is short, and they decay with the sequence length \cite{Amigo2007}. Considering false forbidden permutations, if the selected EEG data length is short, the pTIR of classified sleep EEGs might not be reliable. The distribution of individual permutations (DIPs) is expressed in Eq.~(\ref{eq4}), where $N(\pi_{I})$ and $N(\pi)$ represent the amount of individual permutation $\pi_{I}$ and existing permutation $\pi$, respectively. Taking $m$=5 as an example, the effect of data length on the DIPs of five groups of sleep EEGs is illustrated in Fig.~\ref{fig4}.  

\begin{eqnarray}
	\label{eq4}
	\mbox{DIP} = N(\pi_{I})/N(\pi)
\end{eqnarray}

Fig.~\ref{fig4} shows that sleep EEG signals contain individual permutations whose probability distributions are affected by the signal length. With the increase in data length from 1 to 10 s (2500 points), the DIP values of sleep EEGs decrease stepwise and tend to converge when the data length was larger than 20 s (5000 points). Therefore, when the data length is short, there exist false forbidden as well as false individual permutations, and they decay when the data length increases. The DIP is also related to sleep conditions when the EEG length is more than 20 s. SIII and REM EEGs have larger DIPs when awake and SI EEGs exhibit smaller DIPs. The existence of forbidden and individual permutations suggests that subtraction-based index is a necessity in quantitative TIR. Moreover, the length of EEG signals cannot be too low; they should be greater than 30 s in this study, otherwise these false individual permutations will lead to unreliable pTIR for the nonequilibrium analysis of sleep EEGs.

Overall, there are equal values and individual permutations in sleep EEGs that affect pTIR analysis. Hence, the two basic ordinal patterns must be distinguished considering the necessity of equal-value permutation, the subtraction-based probabilistic difference and requirement of data length also should be paid attentions in pTIR of sleep EEGs.

\subsection{pTIR in sleep EEGs}
\begin{figure*}[htb]
	\centering
	\includegraphics[width=17cm,height=6.4cm]{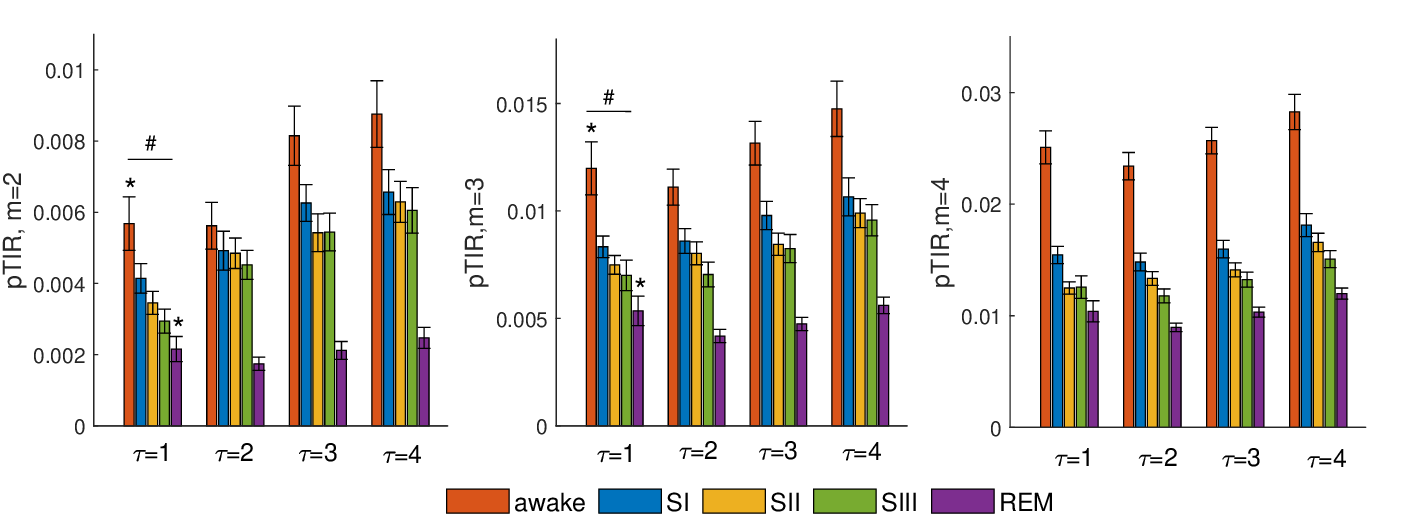}
	\caption{pTIR (mean$\pm$standard error) of wake, SI–SIII, and REM EEGs. \# indicates p$<$0.0001 across all stages using Kruskal--Wallis test and * suggests p$<$0.01 between pTIR of sleep EEGs and others using Mann--Whitney U test.}
	\label{fig5}
\end{figure*}

Given the DIPs in sleep EEGs, dimension is set to 2, 3, and 4 in an enumerative manner. The increase in delay is equivalent to the reduction in signal sampling frequency, i.e., 250/$\tau$ Hz \cite{Yao2021DES,Yao2023tDES}. To satisfy the Nyquist sampling rate (i.e., more than twice the signal band), the delay is set to 1--4, thus maintaining sufficient information about sleep conditions in EEG data. The pTIR of sleep EEGs in awake, SI--SIII, and REM stages are shown in Fig.~\ref{fig5}.

Contrary to the DES in Fig.~\ref{fig3}, the pTIR of sleep EEGs exhibit decreasing trends as the sleep depth increase, as shown in Fig.~\ref{fig5}. As the subjects come into sleep from wakefulness, the pTIR of EEG signals significantly decrease; moreover, as the sleep depth increase, the pTIR consistently decrease. Particularly, when subjects enter the REM state, the pTIR of EEGs show a considerable decrease. The choice of dimension and delay do not affect the trend of EEGs' pTIR with the increasing sleep depth, but it affect the statistical discriminations. According to statistical results, when $m$=3 and $\tau$=1, the pTIR of EEGs in the five sleep stages exhibit optimal classification (p$<$0.0001 for Kruskal--Wallis test). As shown in Fig.~\ref{fig5}, more significant reductions are observed in the pTIR of EEGs from the awake to sleep and SIII to REM stages when $m$=2 and 3 and $\tau$=1. Therefore, time irreversible features of brain electric activity decrease as the sleep stages advance, and more significant differences are observed in nonequilibrium features during the awake--sleep transformation and REM state. For comparison, the probabilistic differences between symmetric AmPs, i.e., pTAS, are also calculated. The pTAS and pTIR yielded the same results, suggesting that equal-value AmPs reliably characterize the temporal structure of vectors in sleep EEGs and are not affected by the reverse process in backward time series.

\begin{figure*}[htb]
	\centering
	\includegraphics[width=17cm,height=11cm]{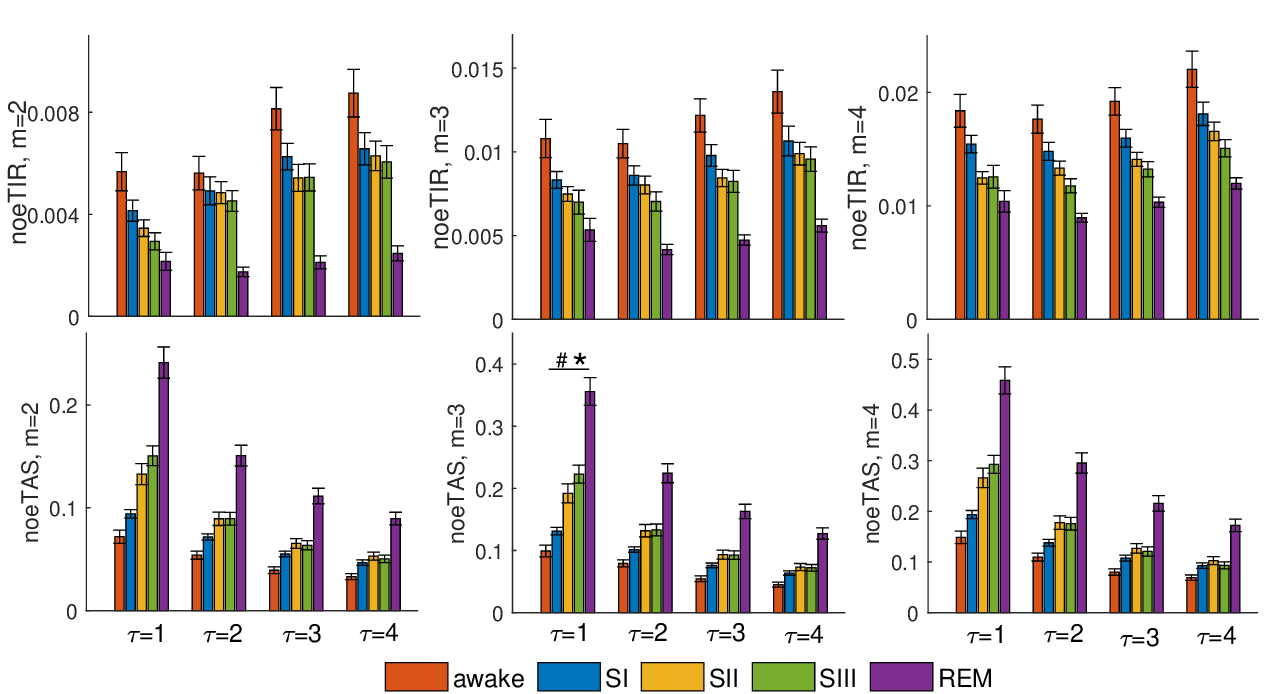}
	\caption{Contradictory noeTIR and noeTAS (mean$\pm$standard error) in wake, SI–SIII, and REM EEGs. \# indicates p$<$0.0001 across all stages using Kruskal--Wallis test and * suggests p$<$0.01 between each of the two stages of sleep EEGs using Mann--Whitney U test.}
	\label{fig6}
\end{figure*}

It should be noted that equal values significantly affect the pTIR, and sleep EEGs generally contain numerous equal values \cite{Yao2023tDES}. Given the amplitude fluctuations measured by DES in Fig.~\ref{fig3}, the noeTIR and noeTAS of EEGs under the five sleep stages are calculated and shown in Fig.~\ref{fig6}.

As evident in Fig.~\ref{fig6}, noeTIR and noeTAS of the five groups of sleep EEGs considerably differed and even showed completely contradictory results based on the nonequal AmP. The comparison shows that noeTIR results exhibited a consistent classification of sleep stages with the pTIR, while noeTAS had contradictory results. Although pTIR and noeTIR demonstrated similar decreasing trends with the increase in sleep depth, they were not the same in these EEGs. Statistically, noeTIR of sleep EEGs was not significantly different, while noeTAS with $m$=3 and $\tau$=1 effectively differed in the five groups of EEGs (p$<$0.01 for Mann--Whitney U test; p$<$0.0001 for Kruskal--Wallis test). Figure~\ref{fig2} shows that when the signal contained equal elements and was reorganized in ascending order, the probabilistic differences of permutations between the forward and backward series were closer to the pTIR, while those of symmetric permutations had a greater deviation. Note that neither noeTIR nor noeTAS yielded correct results. Taking $m$=2 as an example, only up, down, and equal forms are observed, among which, equality was transformed into up in both the forward and backward series. The pTIR, noeTIR, pTAS, and noeTAS were calculated using Eq.~(\ref{eq5}), where $up_{p}$, $down_{p}$, and $equal_{p}$ represent the probability of up, down, and equal values, respectively, and $up_{p}/down_{p}$ denotes the probabilistic difference between $up_{p}$ and $down_{p}$ using $Y_{s}$.

\begin{eqnarray}
	\label{eq5}
	\left\{
	\begin{array}{lr}
		\mbox{pTIR}: 0.5*(up_{p}/down_{p}+down_{p}/up_{p})\\
		\mbox{noeTIR}: 0.5*(\frac{up_{p}+equal_{p}}{down_{p}+equal_{p}}+down_{p}/up_{p})\\
		\mbox{pTAS}: up_{p}/down_{p}\\
		\mbox{noeTAS}: (up_{p}+equal_{p})/down_{p}
	\end{array}
	\right.
\end{eqnarray}

In Eq.~(\ref{eq5}), pTIR is the same as pTAS, but different from noeTIR and noeTAS. Due to equal being mistaken as up, both noeTIR and noeTAS wrongly measure the time reversibility of sleep EEGs. Evidently, if there is no equality, pTIR, noeTIR, pTAS, and noeTAS are the same.

\begin{figure*}[htb]
	\centering
	\includegraphics[width=17cm,height=5.9cm]{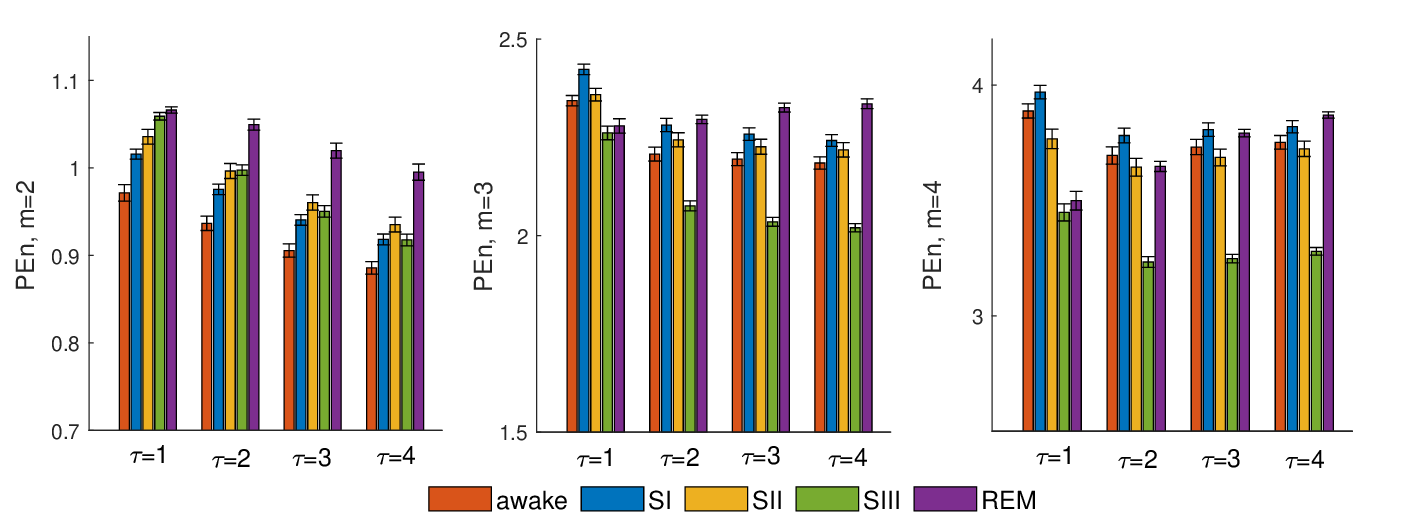}
	\caption{Permutation entropy (mean$\pm$standard error) of wake, SI-SIII, and REM EEGs based on equal-value AmP.}
	\label{fig7}
\end{figure*}

To compare with the pTIR, PEn \cite{Bandt2002,Yao2020APL}, i.e., the Shannon entropy of permutation probability $p(\pi)$, given in Eq.~(\ref{eq6}), is further employed in the analysis of sleep EEGs. It is a widely applied statistical parameter used to measure the nonlinear characteristics of complex systems \cite{Bandt2002}. The selection of OrP and AmP does not affect the calculation of PEn because PEn is the mean information contained in the existing ordinal patterns \cite{Yao2022Perm}, while equal values affect PEn in signal analysis \cite{Bian2012,Zunino2017,Cuesta2018}. PEn based on equal-value AmP with delay from 1 to 4 and dimension ranging from 2 to 4 of sleep EEGs is displayed in Fig.~\ref{fig7}.

\begin{eqnarray}
	\label{eq6}
	\mbox{PEn} = - \sum p(\pi) ln p(\pi)
\end{eqnarray}

As evident from Figs.~\ref{fig7} and \ref{fig5}, PEn is observed to be the opposite of pTIR in the five groups of sleep EEGs. When $m$=2, PEn of EEGs increases as the subjects fell asleep and sleep depth increased, suggesting that EEGs have larger complexity with the increase of sleep stages. As $m$ increases, no consistent trend is observed in the PEn complexity of EEGs with the increasing sleep stages. Statistically, the PEn of five groups of sleep EEGs is not significantly different. Therefore, pTIR is more reliable for sleep stage classification owing to the quantification of EEG nonequilibrium features. 

The contradictory results of pTIR and PEn can be explained by their statistical concepts. Shannon entropy and TIR compute the difference in probability distributions, but they focus on different sets of probabilistic differences. PEn calculates the average amount of information contained in the distribution of permutations, i.e., the probabilistic differences in all existing permutations, whereas pTIR measures the difference between forward and backward permutation series or that between symmetric permutations of a series. Therefore, when the permutation probability difference is smaller, the entropy complexity of signals will be higher, while the TIR will be smaller. This is the fundamental reason for the opposite outcomes of pTIR and PEn in sleep EEGs, which is consistent with the contradictory results obtained in our previous report on heartbeats \cite{Yao2020APL}. Moreover, because they characterize complexity and nonequilibrium features, the results exhibit the diversity of features from different perspectives of complex physiological signals and enable us to explore complex systems more comprehensively. 

\begin{figure*}[htb]
	\centering
	\includegraphics[width=17cm,height=5cm]{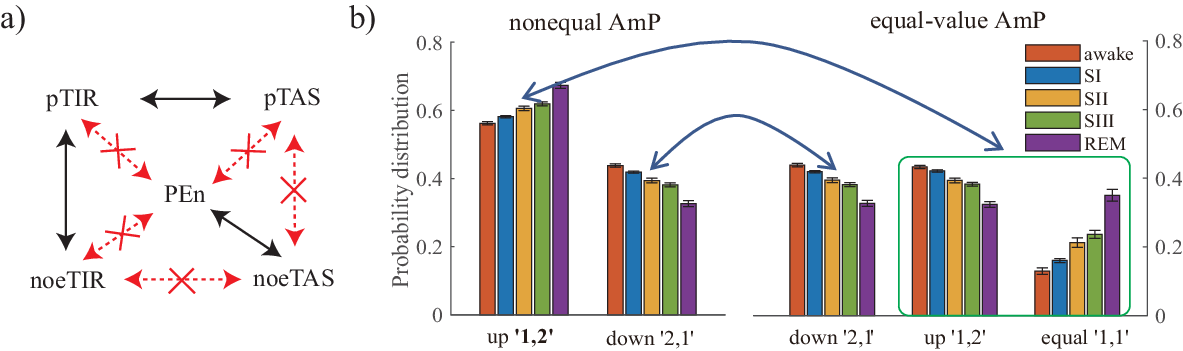}
	\caption{Relationship of permutation analysis in sleep EEGs and probability distributions (mean$\pm$standard error) of AmPs when dimension is 2 and delay is 1. a) Consistent results of pTIR and pTAS are linked by black solid arrows, while contradictory results of noeTIR and noeTAS are connected by crossed red dashed arrows. b) Probability distributions of down `2,1' in the two types of AmPs are the same, while those of bold \textbf{`1,2'} in nonequal AmPs are the sum of those of up `1,2' and equal `1,1' in equal-value AmPs.}
	\label{fig8}
\end{figure*}

The comparative analysis of sleep EEGs demonstrated that the TIR, TAS, and Shannon entropy have a complex relationship, as shown in Fig.~\ref{fig8}. TIR and TAS based equal-value AmP share same results while they yielded contradictory outcomes based on nonequal AmP. The trend of noeTIR is consistent with that of pTIR, while noeTAS exhibited different results in the five groups of sleep EEGs. PEn also presented contradictory findings in comparison to pTIR. To determine the difference between these associated measures and the influence of equal values, probability distributions of AmPs are further analyzed numerically. Taking $m$=2 and $\tau$=1 as an example, the probabilities of AmPs with and without equal values of five groups of sleep EEGs are shown in Fig.~\ref{fig8}.

Figure~\ref{fig8} shows that the distribution of false up (bold \textbf{`1,2'}) in nonequal AmPs is a combination of real up (denoted by `1,2') and equal (denoted by `1,1') in equal-value AmPs because equal values are replaced by up patterns, and that of equal values is the DES of sleep EEGs in Fig.~\ref{fig3}. The distributions of downs (denoted by `2,1') are the same in the two types of ordinal patterns. In nonequal AmP, the probability distribution of up increases with the sleep depth, while that of down presents the opposite trend. In equal-value AmP, the probability distributions of up and down are significantly close, and both decreased as the sleep depth increased. Equal-value AmPs further confirm our conclusion, as shown in Fig.~\ref{fig3}, that fluctuations in the EEG amplitude decreased as the sleep depth increased. Owing to the decrease in amplitude fluctuations in sleep EEGs, the probabilities of both ups and downs in sleep EEGs decrease, while those of equal values (i.e., DES) significantly increase with the sleep depth. 

The probabilistic difference in AmPs and different ways to use them directly influenced the different outcomes of pTIR, pTAS, and PEn in sleep EEGs. According to Eq.~(\ref{eq5}), numerical simulations suggest that the probabilistic difference between false up (\textbf{`1,2'}) and down (`2,1') in nonequal AmPs, i.e., (`1,2'+`1,1')/`2,1', increased with the sleep depth; hence, noeTAS in Fig.~\ref{fig6} also increased with the sleep depth. In the backward EEG signal, equal values were transformed into up `1,2'. Therefore, noeTIR is measured using the sum of forward-backward probabilistic differences between up and down, i.e., 0.5*[(`1,2'+`1,1')/(`2,1'+`1,1')]+`2,1'/`1,2', exhibiting a decreasing trend in sleep EEGs, as shown in Fig.~\ref{fig6}. However, neither noeTIR nor noeTAS truly quantified the nonequilibrium characteristic of sleep EEGs. Meanwhile, in equal-value AmPs, pTIR represented the probabilistic difference between up and down because equality indicated time reversibility and temporal symmetry, while PEn measured the probabilistic difference between up, down, and equality. Numerical results indicated that the probabilistic difference in up and down decreased as the sleep depth increased, while the average amount of information contained in up, down, and equality increased. Therefore, the pTIR in Fig.~\ref{fig5}, i.e., up-down probabilistic differences, decreased with the increasing sleep depth, while the PEn of EEGs in Fig.~\ref{fig7} increased with the sleep depth. Hence, the probability distribution of sleep EEG permutation and different usage of probabilistic difference of pTIR, pTAS, and PEn are responsible for the conflicting results in sleep EEGs.

\begin{figure}[htb]
	\centering
	\includegraphics[width=9cm,height=5.9cm]{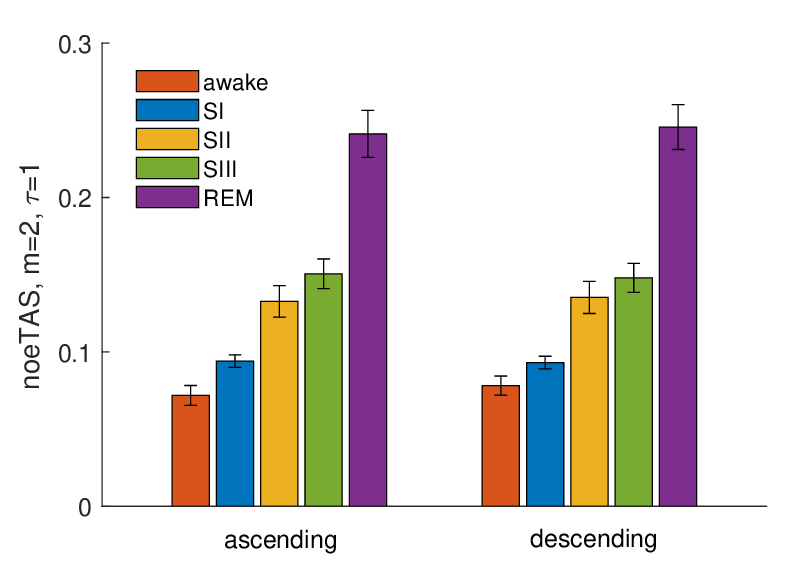}
	\caption{Different noeTAS (mean$\pm$standard error) in wake, SI-SIII, and REM EEG data in ascending and descending ordinal patterns.}
	\label{fig9}
\end{figure}

In ordinal patterns and permutation analysis, elements are reordered in the ascending order; however, it remains unclear how these methods perform of a series reordered in the descending order. In ordinal patterns, if equal values are organized according to their order of occurrence, they will be treated as down in nonequal permutations in the descending order. If indexes of equal values are modified to be the same in their corresponding groups, the AmP and OrP have the same relationship as that shown in Fig.~\ref{fig1}; moreover, the results for TIR, TAS, and PEn show no difference in the ascending order. Otherwise, if the indexes of equal values are not modified, these permutation methods might still generate incorrect results as the ascending order. Particularly, noeTAS yields different outcomes in the ascending and descending orders. The results of noeTAS ($m$=2 and $\tau$=1) for sleep EEGs are shown in Fig.~\ref{fig9}.

In the ascending order, noeTAS is the probabilistic difference between `$up_{p}+equal_{p}$' and `$down_{p}$', while in the descending order, noeTAS is measured by the probabilistic difference between `$down_{p}+equal_{p}$' and `$up_{p}$'. The two noeTAS share consistent results, i.e., their value increases as the sleep stage progresses; meanwhile, they are both incorrect.

Ordinal pattern is a simplified alternative to the vector; their construction is a coarse-grained procedure to signals. Weak noises and artifacts are eliminated during the generation of permutations; therefore, permutation analysis exhibits evident noise insensitivity \cite{Bandt2002,Bandt2016,Bandt2020}. To further test the pTIR, Gaussian noises are added and some frequency components are removed from the sleep EEG data of the five groups. These procedures through software (e.g., MATLAB in this work) eliminate equal values and the information conveyed by DES \cite{Yao2023tDES} from sleep EEG data. Therefore, no difference was observed between TIR and TAS based the ordinal pattern, regardless of the ascending or descending order. Taking $m$=2 and $\tau$=1 as an example, when Gaussian noises with signal noise ratio (SNR) from 0-0.01 dB are added, the results are not significantly affected that the pTIR values of EEG data decrease as the sleep stage progresses. Alternatively, when a 0.3–35 Hz bandpass filter is implemented to the sleep EEG data, pTIR also shows a decreased trend with the increase of sleep stages, confirming the robustness of permutation analysis \cite{Bandt2002,Bandt2016,Bandt2020}.

According to the research on pTIR for sleep classification, pTIR effectively quantifies nonequilibrium characteristics in EEGs, yielding more reliable results than the entropy measure. Owing to the important influence of equal values on ordinal patterns, equal-value permutation is necessary in the pTIR analysis of sleep EEGs. The TIR, TAS, and PEn based on equal-value and nonequal permutations yielded different or even opposite results in the 5 groups of sleep EEG signals; however, this helped us in gaining a deeper understanding of the differences between analytical methods, important role of equal values, and multifaceted characteristics of complex physiological signals.

\section{Discussions}
In the research on pTIR in sleep EEGs, several issues should be further discussed. 

Equal values in time series and their effect on permutation analysis should be given sufficient attention. Equal values are generally ignored considering the limitation resolution of ADC, particularly coarse-grain quantization. In traditional signal processing theory, an arbitrary time series with a weak stationarity exhibits a continuous distribution; therefore, equal values are rare and can be broken numerically by adding small random perturbations \cite{Bandt2002}. The traditional treatment to equality is based on the assumptions that equal values have no significant effect on signal processing and they do not contain information about systems. These assumptions are not correct. Equal values significantly impact permutation analysis. Moreover, they are necessary for the comprehensive construction of ordinal patterns \cite{Yao2022Perm}. As shown in Fig.~\ref{fig1}, irrespective of the OrP or AmP, the structural information of vectors can be fully displayed only when equal values are considered. The distribution of equal values in some signals conveys important information and has significant effects on the construction and probability of ordinal patterns, thus yielding different or even contradictory results, such as the results presented in Figs.~\ref{fig5} and~\ref{fig6}. According to our previous report, PEn and pTIR exhibited contradictory results for PhysioNet heartbeats data with and without equal values \cite{Yao2020APL}. In time reversibility, vectors with equal values might be self-symmetric, such as those in Figs.~\ref{fig1} and~\ref{fig2}, and have definitive physical implication, i.e., time reversibility and temporal symmetry \cite{Yao2022Perm,Yao2023CNS}. Under acceptable ADC resolution, the DES are an effective parameter for quantifying signal amplitude fluctuations in extreme forms, i.e., zero fluctuation. The advantages of DES have been observed in the characterization of sleep and epileptic EEGs \cite{Yao2021DES,Yao2023tDES}; they also exhibit reliable performance in heartrate data (derived from electrocardiography) to reflect the decrease in heart rate variability with age and heart failure \cite{Yao2019E}. If the ADC has a high resolution or the data undergo preprocessing (e.g., software filtering), equal states are rarely observed and amplitude fluctuation information cannon be detected; in this case, a low-pass threshold can be established to filter the differential states for measuring the amplitude fluctuation \cite{Yao2023tDES}. Otherwise, the equalities can be increased by further coarse graining the digital signal, such as the partition symbolic transformation that resembles the ADC. It should be noted that the results of DES were consistent with those of pTIR for sleep EEGs. The amplitude fluctuation of signals is the most direct representation of system information, which is influenced by various factors such as frequency composition and dynamic characteristics. According to Figs.~\ref{fig3} and~\ref{fig8}, the DES of EEG signals increased with the sleep depth, indicating that the amplitude fluctuation decreased. In the previous analysis of heartrate signals, amplitude fluctuations in heartrate decreased with age and heart failure, consistent with the theory of loss of complexity \cite{Yao2019E}. Our results further confirmed the positive correlation between amplitude fluctuations and time irreversible features, and whether there exist other factors closely related to amplitude fluctuations require further research. Overall, the distribution of equal values conveyed important information about amplitude fluctuation and significantly affected signal processing; moreover, it is required to construct reliable ordinal patterns and has explicit physical meaning in TIR; hence, equal values cannon be ignored.

Time reversibility and temporal symmetry are equivalent in statistical definitions, even though they may generate different results in real-world quantification. In pTIR and pTAS, the construction of ordinal patterns, particularly the treatment of equal values, plays an important role. In traditional ordinal patterns, equal values are generally ranked according to their order of emergence. Considering double values, there are three kinds of permutations, namely up, down, and equal. In the traditional ordinal, equal values are neglected and treated as up. As shown in Figs.~\ref{fig2} and~\ref{fig8}, the probabilistic difference between 'up+equal' and down is calculated in noeTAS, while that between ‘up+equal’ and ‘down+equal’ is measures in noeTIR. Numerical calculations demonstrated that the difference in forward-backward permutations was closer to the pTIR than that in symmetric permutations, but both were wrong. The noeTIR and noeTAS were both incorrect and not the same. Taking an extreme example, for a series of all-equal values, the pTAS is 1, indicating the difference between all up and zero down, while the noeTIR is 0 for the forward and backward series (it is the same as the pTIR of 1). If more values are considered, the situation will become more complex.

Next, the relationship between PEn and pTIR requires more discussion. Test results indicated that pTIR enables the more effective classification of sleep stages than PEn, consistent with the results of our previous report on epilepsy EEGs \cite{Yao2020ND}. TIR and Shannon entropy are both statistical parameters for measuring probabilistic differences; they are both widely employed in complex process analysis. Shannon entropy quantifies the static complexity and unpredictability considering the amount of information, i.e., the mean logarithmic calculation of all probabilities of permutations. TIR measures nonequilibrium features considering the sum of probabilistic differences between vectors in forward-backward series or pairs of symmetric vectors. Mathematically, if two permutations have larger probabilistic differences, they convey less information while being more nonequilibrium, thus leading to smaller PEn and bigger pTIR. In the special case where all permutations of symmetric vectors have the same probability distribution, PEn reaches a maximum value while pTIR is 0. In another special case where all permutations are single permutations, pTIR is the maximum 1 while PEn varies
with the difference among probability distributions. This is the fundamental reason for different or even contradictory results in the same process. Similar results have also been reported in heartrate analysis \cite{Yao2020APL}; irrespective of nonequal or equal-value permutation, PEn and pTIR yielded contradictory results. Such discrepant results of pTIR and PEn inspired us to gain a more comprehensive and profound understanding of the characteristics of complex systems from different perspectives.

Forbidden permutation is an important influencing factor that is generally overlooked in the pTIR analysis of real-world signals. The existence of forbidden permutations might generate individual permutations, which have no symmetric form in forward series and de not exist in backward series; moreover, their probabilistic difference is zero or infinite considering division-based parameters, which is not appropriate in quantitative TIR. Associated with forbidden permutations, individual permutations also convey information about sleep EEGs. In Fig.~\ref{fig4}, the DIPs generally increased with the sleep stage when the data length was larger than 20 s. This may be because as the sleep depth increased, amplitude fluctuations, type of temporal structure, and number of ordinal patterns decreased, yielding more forbidden as well as individual permutations. Similar associations have also been reported in epileptic EEG analysis \cite{Yao2020ND,Yao2021DES}. Seizure ictal EEGs exhibit abnormally large amplitude fluctuations and TIR owing to the development of synchronous neuronal firings, while seizure-free postictal brain activity exhibits rather smooth amplitude fluctuations as well as smaller TIR \cite{Yao2021DES}. Consistently, brain electric signals under ictal and postictal states demonstrate larger and smaller DIP values, respectively \cite{Yao2020ND}. In a series of related studies, Amigo et al. found that the existence of forbidden patterns is a feature of chaotic dynamics and can be used to distinguish random from pseudorandom orbit generation \cite{Amigo2006}; subsequently, they identified false forbidden patterns \cite{Amigo2007} and detected determinism in noisy time series based on the properties of topological PEn \cite{Amigo2008}. The decay rate of forbidden permutations \cite{Carpi2010} in stochastic processes has been reported to be associated with their correlation structures; moreover, nonlinearity in time series can be possibly detected by the number of forbidden permutations \cite{Kulp2017,David2020}. Given the systematic information conveyed by forbidden permutations and association of DIPs with sleep stages, individual permutations might also be potentially used for feature detection in complex systems.

\section{Conclusions}
In this study, pTIR in sleep EEG particular the dependence on sleep stage and the effect of equal EEG values are analyzed. Main findings are summarized below:

Permutation TIR is an important measure for the quantification of nonequilibrium EEGs. When symmetric vector differences are used for real-time requirements, AmP is more suitable as a direct alternative to vector. Moreover, equal-value ordinal patterns are required because they construct comprehensive vector structures and self-symmetric vectors convey an important physical implication, i.e., time reversibility.

Brain electrical activity demonstrates nonequilibrium features that are influenced by sleep conditions. EEGs during wakefulness exhibit higher pTIR; during sleep and with the advancement of sleep stages, the pTIR values of EEGs significantly decrease. These findings suggest that when people fall asleep, their brain electric activity contains less nonequilibrium characteristics. The effective classification of sleep EEGs suggested that pTIR could serve as an aid for the expert manual annotation of sleep stages.

When constructing ordinal patterns, if equal values are ordered according to their order of occurrence, the values of noeTIR and noeTAS may be inconsistent, and even if noeTIR is closer to the actual result, they would both be wrong. Moreover, noeTAS performs differently over series reordered in the ascending and descending orders. Therefore, the consideration of equal values is necessary to construct reliable permutations, and it is important to modify the indexes of equal values to same forms in permutation TIR.

Equal values and individual permutations affect the pTIR, but both DES and DIP contain important information about sleep EEGs. The DES characterize amplitude fluctuations in sleep EEGs such that as the sleep stages advance, DES values significantly increase as amplitude fluctuations decrease. The DIP may be related to structures that require further investigation.

Comparative analysis of pTIR and PEn as well as numerical simulations of permutation probability distributions verified the different and even contradictory results of time reversibility and entropy complexity, thus providing us with valuable insights on statistical measures and enabling us to explore complex physiological signals more comprehensively. 

\section{Acknowledgment}
The project is supported by the Natural Science Foundation of Jiangsu Province (Grant No.BK20220383), Natural Science Research of Jiangsu Higher Education Institutions of China (Grant No.22KJB110003), Natural Science Foundation of Nanjing University of Posts and Telecommunications (Grant Nos. NY221142, NY222172), Shanghai Municipal Science and Technology, China Major Project (Grant No. 2018SHZDZX01), Key Laboratory of Computational Neuroscience and Brain-Inspired Intelligence (LCNBI) and ZJLab.

\nocite{*}

\bibliography{mybibfile}% Produces the bibliography via BibTeX.

\end{document}